# Early-life Income Shocks and Old-Age Cause-Specific Mortality

**Hamid NoghaniBehambari[1*]  |  Farzaneh Noghani[2]  |  Nahid Tavassoli[2]**

[1] Texas Tech University, Department of Economics, Lubbock, United States
[2] Texas Tech University, Rawls Business School, Department of Management, Lubbock, United States

**ABSTRACT**

This paper investigates the causal relationship between income shocks during the first years of life and adulthood mortality due to specific causes of death. Using all death records in the United States during 1968-2004 for individuals who were born in the first half of the 20th century, we document a sizable and statistically significant association between income shocks early in life, proxied by GDP per capita fluctuations, and old age cause-specific mortality. Conditional on individual characteristics and controlling for a broad array of current and early-life conditions, we find that a 1 percent decrease in the aggregate business cycle in the year of birth is associated with 2.2, 2.3, 3.1, 3.7, 0.9, and 2.1 percent increase in the likelihood of mortality in old ages due to malignant neoplasms, Diabetes Mellitus, cardiovascular diseases, Influenza, chronic respiratory diseases, and all other diseases, respectively.

## INTRODUCTION

The disturbances in the accumulation of health endowments during the early-life period could alter the trajectory of individuals' outcomes later in life. Several strands of empirical economic research have documented the mechanisms through which initial health endowments can explain, partly, the variations of short, medium, and long term individual outcomes including infant mortality, toddler mortality, cognitive development, test scores, completed education, employment, earnings, mortality rates during adulthood, and hazards of old-age cause-specific death (for a recent review, refer to Costa (2015)). According to the fetal origin hypothesis, a nutritional shock during pregnancy can be compensated by an adjustment in vital organs and generally human system development with the main purpose of embryo's survival (Barker, 1990). The lower quality of health endowment of the survived infants provides a channel to later-life outcomes. Another possible explanation for this long-term link is that lower quality early-life conditions may posit individuals in inimical life trajectories (Yeung et al., 2014). Being born in a poor family might leave individuals with adverse health endowments but also could lead to lower education, lower likelihood of employment, and lower earnings. The lower earnings during adulthood could cause lower access to health care and so an increase in mortality. This effect reinforces the negative relationship between early-life conditions and adult mortality. Another channel is families' selection for educational investment in their healthier/unhealthier children. As stated in Costa (2015), since the marginal benefit of education rises at higher levels of schooling, householdds tend to sort their healthier children into high human capital development. Therefore, healthier children benefit from higher educations and so higher earnings during adulthood. The higher earnings mean higher current nutrition and health-care access and so lower likelihood of mortality. Exploring the long-term health effects of early-life conditions can have important policy implications. It could help policymakers identify the most vulnerable groups in the society, design more accurate health-care policies to increase the health of the society, and decrease the mortality gap among different cohorts and gender-race groups. More importantly, health, as one of the fundamental components of human capital, has been shown to have sizable and statistically significant effects on aggregate output (Bloom et al., 2004; Fogel, 1994; Grimm, 2011). Therefore, a well-designed policy could not only lessen inequality but also contribute to long-term economic growth. This paper explores the effect of a negative income shock during the early-life period on old-age adult mortality due to specific causes of death. The main challenge to build a causal relationship is to find an exogenous shock that had an effect on families' income but is not correlated with their characteristics. Moreover, such shocks should not directly affect individuals' outcomes later in life. A small branch of the literature uses booms and busts in aggregate output as the exogenous source (den Berg et al., 2006, 2011; Yeung et al., 2014). Following this literature, we use business cycle fluctuations at the year of birth and first year of life as the source of shock to households' income. We use Cause-Specific Mortality data, which covers all death records in the United States. For individuals born between 1910 till 1950, and who died between 1968-2004, we find a sizable, statistically, and economically significant effect of early-life fluctuations in real gross domestic product per capita on the risks of mortality due to different causes of death. A 1 percent reduction in GDP per capita at the year of birth is associated with an increase in the likelihood of death due to malignant neoplasms, Diabetes Mellitus, cardiovascular diseases, Influenza, chronic respiratory diseases, and all other diseases by 2.2, 2.3, 3.1, 3.7, 0.9, and 2.1 percent, respectively. The marginal effects are still significant and in some cases larger using business fluctuations in the first year of childhood. In order to explore the potentially different effects of macroeconomic conditions across cohorts, we split the sample into four cohorts based on their year of birth. Each subsample contains at least one severe recession. The negative effect of the recessions has mitigated over time and became even positive for cohorts born between the years 1943-1950. The main reason behind this sign reversion could be the large government anti-poverty policies such as the Emergency Relief Act of 1933, Aid to Families with Dependent Children Act of 1940, and Social Security Act of 1935. Note that this fact is consistent with the literature that mortality in wealthy countries like the US is procyclical (G. Miller and Urdinola, 2010; Ruhm, 2000, 2015). However, these analyses work with datasets that cover the years after the start of these social programs. In the main results, we include dummies to capture the effect of these policies. The negative and statistically significant coefficients confirm the fact that this sign reversion for recent cohorts is, possibly, due to governmental policies to provide a minimum level of welfare (like nutritional sufficiency) for American families. Although there is no established causal effect of government policies on later-life health outcomes, the cohort analysis points to a plausible avenue for future research. Next, we stratify the sample based on race and gender. In all cases, the negative effects of recessions on cause-specific mortality are confirmed. The gender difference varies across models and depends on race and cause of death. However, comparing the marginal effects of whites and blacks, an income shock during childhood is more effective among whites. Higher rates of infant mortality among blacks could partly explain this gap since they let fitter and healthier black individuals survive into adulthood, which shows lower cause-specific mortality later in life. Even after controlling for education, occupation, and marital status, this gap prevails. However, all marginal effects are statistically significant at conventional levels. This paper makes a number of contributions to the current literature. First, while there is limited literature that explores the economic conditions early in life and old-age mortality, no study has investigated this link for US data. More noticeably, no study has investigated early-life conditions and cause-specific death using US data. Therefore, the main contribution of this paper to the literature is that it is the first study to fill this gap in the literature. Second, it contributes to the literature that explores the latent outcomes of early-life negative shocks by exploring the different causes of death among cohorts born in different economic circumstances. Third, by presenting separate sets of marginal effects for different cohorts, this paper offers a plausible mechanism through which recessions started to be a healthy period in the US, which is the starting point of mass government aids and social security programs. Third, using all death records in the US, which provides millions of observations, makes the results of this study more reliable compared to previous studies. The rest of the paper is organized as follows: In section 2, we provide a brief literature review. Section 3 describes the core dataset, sample selection strategy, and gives a brief summary statistics of the final dataset. The identification strategy and main results are discussed in section 4. In section 5, we go over some robustness checks and show the heterogeneity of the coefficients in different sub-samples. Possible issues and drawbacks of the analysis, some concluding remarks, and suggestions for future research are presented in section 6.

---

[*] Corresponding author, e-mail: Hamid.NoghaniBehambari@ttu.edu

**A BRIEF LITERATURE REVIEW**

Recessions could affect the health endowment of individuals through different channels. As the main channel, it affects the wealth and earnings of households. Lower earnings could be translated into lower nutritional access or health care affordability. This link between economic status early in life and short and long-term health outcomes is well established in the literature. Ferrie and Rolf (2011) use longitudinal data which covers the years 1895-2005 and show that US-born male who lived in a low socioeconomic household before age 5 are more likely to die younger, conditional on survivor until age 70, and more likely to die from heart diseases than those individuals who were in high-SES households before age 5. Using Panel Study of Income Dynamics (PSID), Hoynes et al. (2016)investigate the long-run impacts of participation in the Food Stamp Program. Access to food stamps during early childhood is associated with lower odds of metabolic syndrome and raises the economic self-sufficiency of women in adulthood. The important implication of their paper is the causal effect of a direct policy-driven channel, namely nutrition, during childhood on health outcomes later in life. A small strand of literature uses fasting during the Islamic holy month of Ramadan as a plausibly exogenous nutritional shock in utero and investigates its impacts later in life. Being exposed to Ramadan during prenatal development is associated with lower birth weight, reduces the number of male births, increases the likelihood of disability during adulthood (Almond and Mazumder, 2011), lowers the test scores at age 7 (Almond et al., 2014), intensifies the symptoms indicative of coronary heart problems and type 2 diabetes during old ages (Van Ewijk, 2011), diminishes cognitive and math test scores at school and decreases working hours during adulthood (Majid, 2015). Based on these studies, we expect that a negative income shock caused by a fall in GDP has adverse health effects on children in utero or in their early years of life by reducing nutritional stock available to households. Another possible mechanism through which a recession could influence health outcomes is the reduction in pollution. Recession comoves strongly with pollution and pollutant industries including the manufacturing sector (Eaton et al., 2016; Tavassoli et al., 2020). It is now well established that exposure to pollution during the antenatal development period has adverse health impacts for newborns. Isen et al. (2017) use US administrative data to evaluate the long-term effects of the 1970 Clean Air Amendment Act. They exploit the variation of differential exposures of counties that were affected by the CAAA and find that higher exposure in the year of birth is associated with lower earnings and labor force participation at the age of 30. Chay and Greenstone (2003) use variation in pollution exposure in US counties caused by 1981-1982 recession to estimate the effects of pollution on infant mortality rates and find that 1 percent reduction in Total Suspended Particulates (TSP) will cause a 0.35 percent decline in infant mortality. Through this channel, a positive relationship is expected between recession and health. Following the influential paper of Ruhm (2000), this theory was supported by empirical evidence of procyclical fluctuations in total mortality and several cause-specific death (Coile et al., 2014; Janet Currie et al., 2015; McInerney and Mellor, 2012; D. L. Miller et al., 2009; Ruhm, 2015). In addition, an economic collapse during a severe recession could force mental pressure on pregnant mothers and cause more stressful pregnancies. This pathway, although not widely explored, has been shown to increase the probability of a low birth weight child and reduce the sex ratio (Olafsson, 2016). The effect of these shocks could be deteriorated by parental over- or under-investment in children. The parental investment could reduce the health endowment gap between their offspring or reinforce this gap if they decide to invest more in their healthier children (Janet Currie, 2011; Frijters et al., 2010, 2013; Yi et al., 2015). Restrepo (2016) finds evidence that low educated parents allocate more resources to their offspring who had initially normal birth weight compared to their low birth weight children while high-educated mothers compensate this health gap among their children. Mortality rates during adulthood and old-age could be partly explained by early-life nutritional and economic conditions. Yeung et al. (2014) explore this path and find that an adverse income shock caused by a recession during pregnancy and the first year of life will increase the risks of old-age cause-specific mortality. It increases the probability of death due to cancer among males and females by about 8 and 6 percent, respectively. In addition, it increases female mortality due to cardiovascular diseases by roughly 5 percent. Den Berg et al. (2006) use a longitudinal dataset covering the years 1812-2000 for individuals born in the period 1812-1912 in Netherlands and document that household economic conditions early in life can explain adult mortality rates. They exploit booms and busts during childhood as an instrument for individual economic conditions. Applying a hazard analysis, they find that being exposed to a boom early in life is associated with a 9 percent reduction in adult mortality rates. Using the same instrument, Van den Berg et al. (2011) show that being born under a recession increases the probability of cardiovascular mortality later in life. Den Berg and Gupta (2015) introduce a causal pathway for this effect. They show that marital status, as a determinant of adult mortality, is itself affected by economic conditions early in life. Among women, longevity is reduced upon marriage in case they are born under adverse economic conditions. However, marriage has a protective effect for men. Married men enjoy longevity and the marital status does not depend on economic conditions in early-life. On the other hand, male mortality rates comove, negatively, with business fluctuations in their early childhood. Other relevant papers confirm the influence of economic conditions during early-life and later outcomes including health, mortality rates, and old-age cause-specific mortality rates (Banerjee et al., 2010; Case et al., 2005; D. M. Cutler et al., 2007; Flores and Kalwij, 2014; Frijters et al., 2010; Myrskylä, 2010a, 2010b; Myrskylä et al., 2013; Noghanibehambari et al., 2020; Rao, 2016; Strand and Kunst, 2006).

**DATA AND SAMPLE SELECTION STRATEGY**

The main source of data is the Multiple Cause of Death data files from US Vital Statistics. Since there are multiple interactions between early-life circumstances and individuals' state of birth, we restrict the sample to the years 1979-2004, which contain the information on the state of birth. All individuals who were not born in the US excluding its territories are eliminated. The main filter for the final dataset is to choose a time-window for individuals' year of birth. Going further back in time has the advantage of larger sample sizes and including more cohorts. The problem is the selection bias due to earlier mortality rates. For instance, cohorts born between 1900-1910 who died between the years 1979-2004 have a minimum and maximum age range of 69 and 104. Old-age death due to different causes could have started earlier during their 50s or 60s. Thus, the estimations will probably suffer from sample selection. To alleviate this problem, we let the maximum age at the beginning of the observation-window to be 69. This means that the year of birth is restricted to be above 1910. Moreover, since the focus of this study is old-age mortality, individuals below age 55 are excluded. Therefore, all death records in the final dataset have occurred to persons who were born between 1910-1950, a period with several severe recessions and significant booms in the US economy. Ultimately, in order to overcome the processing power limitation due to the large sample size, we use a 20 percent random sample withdrawn from the final dataset in all analyses. Since the sample size still contains millions of observations, we expect to see similar results. Indeed, when we use a 1 percent, 5 percent, and even 30 percent random sample, the marginal effects and their standard errors are quite similar to the current level of random sampling. We do not restrict this random sampling to be from a specific subpopulation or over- or under-representing some subsamples. The random sample is drawn from the full sample. This random sampling has been done in the previous literature (J. Currie and Moretti, 2003). Table 1 depicts a cross-tabulation of causes of death records by year of birth by gender. Cardiovascular diseases are the main causes of death for both genders and malignant neoplasm diseases are the second. In most cells, male death rates are higher than that of females. Historical estimates of US GDP per capita are extracted from Jordà et al. (2017). The increasing trend in GDP per capita during 20th century comoves with other social and noneconomic outcomes like provision of public health, health care availability, progress in sewage systems, water cleaning, persistent increase in the number of schools and universities, and more noticeably improvements in governmental welfare policies. These secular trends make the use of GDP per capita as the instrument of economic conditions inappropriate. Therefore, following Yeung et al. (2014), we decompose the time series into a time trend and a cyclical component using the Hodrick-Prescott (HP) filter with a smoothing parameter of 200 (Hodrick and Prescott, 1997). Figure 3 illustrates the trend and fluctuations for real GDP per capita between the years 1910-1950. Starting around 1914, there are several recessions during this period (including the big recession) and several considerable boom periods (including World War II years). Note that a positive deviation from the trend refers to a boom in the economy and a downturn from the trend points to a recession. Figure 4 shows a descriptive effect of being born during a recession (boom) on the average age of death from cardiovascular diseases. Visually, except for individuals who died in their late 70s, being born during a recession increases the rates of death from cardiovascular diseases at each given age. One possible drawback in assessing the latent effects is the selection of most fit individuals to reach maturity. Fetal deaths and infant mortality in periods of hardship can let stronger newborns survive. This fact causes the real effects of early-life conditions on adult outcomes to be underestimated. To alleviate this issue we use two other variables that can partly control for secular trends in the health of infants and other macro conditions. First, we use historical average grain prices interacted with state of birth. Using historical meat prices reveals very similar estimates. This dataset is extracted from Jacks (2013) which covers real prices (1900=100) of barley, corn, rice, rye, wheat, beef, hides, lamb, and pork among other commodities in the years 1850-2017. To proxy for the health environment during childhood, we use childhood mortality rates at the year of birth. The time series of childhood mortality rates for children 1-4 years old (interacted with state of birth in all specifications) are extracted from the National Centre for Health Statistics. Table 2 provides summary statistics of the final dataset. Only over a few years, Vital Statistics asked about education and occupation. Over the years 1989-1999, information on both variables is available. This table covers the selected years in order to provide information on education, too. For all causes of death, the median age at death is higher if individuals are born during a boom compared to those born during a bust. Born during either a recession or a boom, females occupy a lower portion of each cause-specific death. As expected, log average grain prices are lower during busts and higher during booms.

**Additional variables**

During this period, the main inflection point in American households' welfare, especially during recessions, was the start of a series of government social welfare programs. We define three indicators to account for the three most important of these acts: Emergency Relief Act commencement (1933), Aid to Families with Dependent Children (AFDC) that was enacted in 1940, and Social Security Act (1935). Moreover, we include in all formulations a dummy to control for the 1918 influenza pandemic which has been shown to have large health impacts among other outcomes during adulthood and old age (Almond, 2006; Almond and Mazumder, 2005). Dustbowl in the Great Plains during the 1930s was another possible channel to affect health since it damaged topsoils and agricultural products in certain ecological places. We include dummies to capture this phenomenon for states (states of birth) that were mostly affected: Texas, New Mexico, Oklahoma, Kansas, and Colorado. In addition, we use 5-year dummies to account for all other secular trends in health improvements and other macro influences that changed over time but affected all individuals uniformly.

**METHODOLOGY**

We take two approaches to reveal the causal relationship between early-life conditions and cause-specific mortality: nonparametric analysis in which we draw Kaplan-Meier survival curves to explore an illustrative relationship, and parametric analysis in which we apply extended Cox model for hazard analysis. Two periods in early-life are the main interest in this study: year of birth and first year of life. Vital Statistics does not ask about the month of birth. Therefore, year of birth could attribute only to the period in utero or only the first year of life or both. This fact should be considered while interpreting the coefficients of the year of birth (YOB) as the variable of interest. Hence, (YOB+1) refers to either the first or second year of life. However, we report all coefficients separately in all specifications.

**Non-parametric analysis**

Figure 1 and Figure 2 depict the Kaplan-Meier survival curves for different causes of death in different stages of life for cohorts born during an expansion versus cohorts born during a recession. The sample is stratified based on gender. It is well established that there is a gender difference in mortality and cause-specific mortality. Figure 1 shows the survival curves of males and Figure 2 depicts that of females. We selected two subsequent periods of expansion and recession during 1912-1913 (boom) and 1914-1915 (bust period) for three reasons. First, there are no government interventions during this period to deteriorate the income shock to households. Second, neither the boom nor the bust is as severe as other recessions in the years 1910-1950. Obtaining the latent effects of early-life conditions during a mild boom-bust period suggests similar results for more severe business fluctuations. Third, public health improvements increased over time and so act at the benefit of later cohorts who were born during recessions. This will cause an underestimation of true effects if we compare two subsequent cohorts born during expansion-recession while leads to an overestimate if we compare cohorts born during succeeding recession-expansion periods. The latent effects of born in a recession are more informative at older ages. For any level of mortality rate, being born in an expansion is associated with higher age at death. Cohorts who were born during a recession are more probable to die younger considering each cause of death. Referring to the upper left panel of Figure 1 and Figure 2, the differential effect begins at age 70 for deaths due to malignant neoplasm diseases. There is no obvious difference between the start of this divergence among males and females. The middle left panel of both figures shows the survival curves for death due to cardiovascular diseases. The survival of both cohorts starts to diverge considerably around age 80, for both males and females. This fact reveals that, at least descriptively, the effects of adverse conditions in childhood can remain passive until much older ages.

**Parametric analysis**

Following the literature (Myrskylä et al., 2013; Strand and Kunst, 2006; Yeung et al., 2014), we assume that conditional on individual characteristics the causes of death are independent from each other. we apply extended Cox model which allows for time dependent variables (Therneau and Grambsch, 2013). A simple Cox model can be written in the following form:

$$h(t, x(t)) = h_0(t) e^{\sum_{i=1}^{p} \beta_j x_j} \qquad (1)$$

Where t is the survival time, h is a hazard function, x represents covariates, and h_0 is the baseline hazard. The main results are presented in Table 3. Each column represents a separate regression. Individual characteristics include age dummies, fixed effects for the current state of residence, gender, and race. However, the literature has documented a differential path for mortality among different races and genders. Including a dummy for the race in these models will shift the survival trajectory vertically. However, stratifying the data will allow each stratum to form its own trajectory, which provides flexibility in survival paths and increases the precision of the estimates. Therefore, instead of adding a set of dummies for the state of birth, state of death, race, and gender as control variables, we stratify the model by these variables. Other control variables are explained in section 3.1. As shown in the first column of each panel in Table 3, a 1 percent reduction in the cyclical component of GDP per capita during the year of birth is associated with 3.2, 2.8, 2.6, 1.9, and 3.7 percent increase in the likelihood of mortality due to malignant neoplasm diseases, diabetes mellitus, cardiovascular diseases, Influenza and pneumonia diseases, and chronic respiratory diseases, respectively. The effects are larger for exposure to business fluctuation in the first year of life. Next, to account for the secular increase in health conditions during childhood, we add an interaction of state of birth with rates of childhood mortality. To account for other macro conditions, we add the log of average grain prices interacted with the state of birth. The results are reported in Table 4. The magnitude of the estimates falls but remains significant at 1 percent level. In the next stage, we add education, marital status, and a set of dummies for occupation. Due to data limitations explained in section 3, the observations fall in the years 1989-1999 only. The results of full specification models for different causes of death are reported in Table 5. A 1 percent decrease in the cyclical component of real GDP per capita during the year of birth is associated with 2.2, 2.3, 3.1, 3.7, and 0.9 percent rise in the probability of death due to malignant neoplasm diseases, diabetes mellitus, cardiovascular diseases, Influenza and pneumonia diseases, and chronic respiratory diseases, respectively. The estimates are quite similar to the results of Table 4 which implies that education and type of occupation do not mitigate the latent effects of adverse conditions in the early-life. The marginal effects of the first year in life are larger than coefficients of the year at birth for most causes of death in all three tables. Among the causes that we explore here, cardiovascular diseases and chronic respiratory diseases are among the main mortality causes that are linked to adverse conditions during pregnancy and early years of life. For example, Lawlor et al. (2006) investigate the socioeconomic conditions during childhood and cause-specific death at older ages in Sweden and find a positive association between economic conditions early in life and higher risks of death due to respiratory diseases. Using data of the Netherlands, Yeung et al. (2014) find a positive association between economic conditions early in life and hazard of death due to respiratory and cardiovascular diseases.

**ROBUSTNESS CHECKS**

**Race-gender decomposition**

There is evidence on racial and gender gap in mortality (Montez et al., 2011; Satcher et al., 2005). If cause-specific mortality can be partly explained by childhood conditions then the racial and gender gap could provide different pathways from childhood condition into adult mortality and morbidity. To check for this differential effect, we split the sample into four groups: white males, white females, black males, and black females. The full specification models of the group decomposition are reported in Table 6 through Table 8 for different causes of death, separately. The gender difference depends on race and cause of death. Childhood conditions are more effective on while males rather than white females for death due to malignant neoplasm diseases while for diabetes mellitus the impacts are more pronounced for females. Black females are less affected by childhood conditions based on death due to chronic respiratory diseases compared to black males while in other causes of death their respective coefficients are larger. In general, cause-specific deaths in white people are more influenced by their childhood conditions compared to their black counterparts. The latter findings could be partly explained by higher fetal death and infant mortality among black people (Elder et al., 2016). During hardships, infant mortality and fetal death increase among blacks more than whites. Therefore, those black newborns who survive their infancy are the fittest and healthiest and have higher initial endowments. This higher level of initial health increases their life expectancy and reduces the mortality rates at older ages. Thus, the lower coefficients are only a reflection of initial selection rather than the lower susceptibility of blacks to income shocks.

**Cohort decomposition**

Secular improvements in the health environment, changes in the provision of public health, the introduction of new vaccines, eradicating many formerly deadly diseases, and more noticeably the change in governmental policies to neutralize the adverse economic shocks lead to substantial variations in cohorts' health quality (Costa, 2015; D. Cutler and Miller, 2005; Noghanibehambari et al., 2020; NoghaniBehambari et al., 2020; Noghanibehambari and Salari, 2020). To account for this possible heterogeneity, we apply the baseline model separately for different cohorts. The results are depicted in Table 9 through Table 11 for six causes of death and four cohorts. Two criteria are used to make cohort boundaries. First, a range should have at least one major recession and one major boom so that the model can compare different cohorts within our cohort-groups. Second, it should reflect, partly, the medical time-line in the first half of 20th century. Considering these facts, four cohorts based on year of birth are defined, as follows: 1910-1918, 1919-1928, 1929-1942, and 1943-1950. As shown in Table 9, a 1 percent reduction in aggregate business fluctuation is associated with 1.5 and 0.8 percent increase in the likelihood of death due to malignant neoplasm and cardiovascular diseases for cohorts born between the years 1910-1918. Surprisingly, the marginal effects increase to approximately 15 percent for cohorts born in 1919-1928. The signs reverse for cohorts born in the period 1943-150 which implies that recessions are good for the health of newborns and reduces the probability of cause-specific mortality in old ages. This fact is confirmed for other causes of death shown in Table 10 and Table 11. For chronic respiratory diseases, the positive effect of recessions starts at the cohorts born in 1929-1942. This finding is consistent with the literature that recessions could improve the health of newborns (Chay and Greenstone, 2003; Miller et al., 2009; Page et al., 2017; Tapia Granados and Ionides, 2017). However, these studies use the most recent data for the US and more specifically in the second half of the 20th century. In a similar analysis, Cutler et al. (2007) use the Dust Bowl in Great Plains during the 1930s as the exogenous shock to income in utero and investigates its effects later in life. It finds no evidence on such effects on chronic diseases, disability, and infant mortality. One possible explanation for these findings is the effectiveness of welfare programs. Recall that a series of government policies started during the 1930s. These policies, like the Social Security Act, could have provided pregnant mothers with sufficient nutrition. Meanwhile, a recession could lower pollution and supply a cleaner environment. As suggested by Ruhm (2000), recessions could lower obesity, increase physical activity, and decrease habits like smoking and drinking. Therefore, an effective welfare program can offset the effect of recession by extending the access of households to nutritional resources that is the main channel through which adverse economic conditions can affect health. Thus, the positive effects dominate the negative ones.

**DISCUSSION AND CONCLUSION**

The long term relationship between early-life economic conditions and old-age cause-specific mortality has been investigated in several studies using European data (Bhalotra et al., 2017; den Berg et al., 2006, 2011; Yeung et al., 2014). Virtually no study explored this topic using US data mostly due to lack of historical longitudinal data (Costa, 2015; Ferrie and Rolf, 2011). This paper attempted to investigate this question using all death records in US Vital Statistics. It uses the fluctuations in the business cycle at birth as the instrument for economic conditions during infancy and finds that being born during a recession has a significant and substantive effect on the probability of cause-specific mortality. This negative relationship is stronger for older cohorts. In the nonparametric analysis, we showed that the effects on death due to cardiovascular diseases remain latent roughly until age 80. In a full specification of Cox models, we find that 1 percent decrease in the cyclical component of GDP per capita is associated with 2.2, 2.3, 3.1, 3.7, 0.9, and 2.1 percent increase in the likelihood of mortality in old ages due to malignant neoplasms, Diabetes Mellitus, cardiovascular, Influenza, chronic respiratory diseases, and all other diseases. Cohorts born during a recession in earlier periods, e.g. 1910-1930, are more prone to be affected by the economic conditions early in life while the sign of the marginal effects reverses for recent cohorts, e.g. 1940-1950. This cohort analysis suggests that government welfare programs could neutralize the effect of income shock by granting American families a minimum level of nutritional requirements. This could let other positive effects of recessions to dominate the negative effect of an income shock (D. L. Miller et al., 2009). This study has some drawbacks resolving which could point to future research avenues. First, the path from childhood economic conditions to old-age mortality passes through some mediatory channels. Although we tried to proxy contemporaneous socioeconomic status by education, marital status, and occupation, no data is available to control for earnings, retirement age, years of labor force participation, wealth, insurance, and access to health care. All these variables can explain a significant portion of variation in mortality (Gerdtham and Johannesson, 2004). Second, income shocks can adversely affect health during early-life in several ways including lack of sufficient nutrition, stressful pregnancies, lowering access to health care, and decreasing prenatal visits. Disentangling the different mechanisms requires a different strategy. Third, assuming that the main channel is nutritional deficiency, not all households are affected by a recession to the same extent. More educated mothers with higher socioeconomic status are less affected compared to families with lower socioeconomic status. This heterogeneity among individuals from different families is not captured in the current study. Fourth, male embryos are more susceptible to external stressors and so during times of hardships sex ratio decreases. This will also lead to higher fetal death for males. Thus, those male newborns that are born during harsh periods have been fitter, stronger, and healthier in the first place. Hence, male cohorts born during a recession could be, on average, healthier than their female counterparts and possibly healthier than males born during a proceeding expansion. This fact will attenuate the estimation of the impact of childhood conditions on old-age cause-specific mortality

**TABLES**

**Table 1.** Tabulation of Causes of Death by Year of Birth and Gender

|  | Year of Birth: 1910-1920 |  |  | Year of Birth: 1920-1930 |  |  | Year of Birth: 1930-1940 |  |  | Year of Birth: 1940-1950 |  |  |
|---|---|---|---|---|---|---|---|---|---|---|---|---|
|  | Male | Female | Total | Male | Female | Total | Male | Female | Total | Male | Female | Total |
| Malignant Neoplasm | 26.2 | 24.1 | 25.2 | 31.4 | 34.5 | 32.8 | 34.1 | 42.2 | 37.5 | 30.6 | 44.7 | 36.2 |
| Diabetes Mellitus | 2.0 | 3.0 | 2.5 | 2.5 | 3.6 | 3.0 | 2.9 | 3.9 | 3.3 | 3.3 | 3.8 | 3.5 |
| Cardiovascular Disease | 49.5 | 48.1 | 48.8 | 44.1 | 37.4 | 41.2 | 40.5 | 29.9 | 36.1 | 36.8 | 24.9 | 32.1 |
| Influenza and Pneumonia | 2.5 | 2.2 | 2.4 | 1.6 | 1.3 | 1.5 | 1.2 | 1.1 | 1.2 | 1.3 | 1.2 | 1.3 |
| Chronic Lower Respiratory | 4.4 | 3.2 | 3.8 | 3.3 | 3.3 | 3.3 | 2.0 | 2.5 | 2.2 | 0.9 | 1.4 | 1.1 |
| All Other Diseases | 15.4 | 19.3 | 17.3 | 17.1 | 19.9 | 18.3 | 19.2 | 20.3 | 19.7 | 27.1 | 23.9 | 25.9 |
| Total | 100.0 | 100.0 | 100.0 | 100.0 | 100.0 | 100.0 | 100.0 | 100.0 | 100.0 | 100.0 | 100.0 | 100.0 |
|  | (4,399,039) | (4,078,241) | (8,477,280) | (3,474,810) | (2,653,606) | (6,128,416) | (1,674,667) | (1,165,384) | (2,840,051) | (904,704) | (596,554) | (1,501,258) |
| Number of Cases |  |  | 7,477,280 |  |  | 6,128,416 |  |  | 2,840,051 |  |  | 1,501,258 |

*Notes. Data are extracted from Vital Statistics data files. 50% random sample is used.*

**Table 2.** Sample Characteristics by Cause of Death by Recession Indicator at Year of Birth

|  | Malignant Neoplasm |  | Diabetes Mellitus |  | Cardiovascular |  | Influenza and Pneumonia |  | Chronic Respiratory |  | All Other Diseases |  |
|---|---|---|---|---|---|---|---|---|---|---|---|---|
|  | Boom | Bust | Boom | Bust | Boom | Bust | Boom | Bust | Boom | Bust | Boom | Bust |
| Median age at death | 70.37 | 67.77 | 71.65 | 69.31 | 73.33 | 71.50 | 75.60 | 74.26 | 73.33 | 72.33 | 72.60 | 69.38 |
| % | (8.879) | (10.62) | (8.877) | (10.67) | (8.851) | (10.62) | (8.377) | (10.19) | (7.599) | (9.141) | (10.25) | (12.54) |
| Edu: High School | 49.41 | 49.57 | 48.56 | 49.28 | 47.90 | 48.09 | 46.16 | 46.37 | 50.54 | 50.86 | 47.65 | 47.55 |
| % | (50.00) | (50.00) | (49.98) | (50.00) | (49.96) | (49.96) | (49.85) | (49.87) | (50.00) | (49.99) | (49.94) | (49.94) |
| Edu: < High School | 16.48 | 15.36 | 23.18 | 21.80 | 20.49 | 19.57 | 22.19 | 21.71 | 20.52 | 20.11 | 19.11 | 17.73 |
| % | (37.10) | (36.06) | (42.20) | (41.29) | (40.36) | (39.67) | (41.55) | (41.23) | (40.38) | (40.08) | (39.32) | (38.20) |
| Married | 57.54 | 58.57 | 47.83 | 49.00 | 49.19 | 49.92 | 42.06 | 41.64 | 46.33 | 46.02 | 45.30 | 44.56 |
| % | (49.43) | (49.26) | (49.95) | (49.99) | (49.99) | (50.00) | (49.37) | (49.30) | (49.87) | (49.84) | (49.87) | (49.70) |
| White | 87.68 | 87.52 | 80.00 | 79.00 | 86.83 | 85.79 | 87.68 | 86.62 | 92.93 | 92.24 | 86.70 | 84.65 |
| % | (32.87) | (34.15) | (40.00) | (40.73) | (33.82) | (34.91) | (32.86) | (34.05) | (25.64) | (26.75) | (33.96) | (36.05) |
| Female | 45.81 | 46.31 | 54.32 | 52.88 | 45.50 | 44.25 | 46.81 | 46.51 | 45.83 | 46.49 | 48.67 | 46.32 |
| % | (49.82) | (49.86) | (49.81) | (49.92) | (49.80) | (49.67) | (49.90) | (49.88) | (49.83) | (49.88) | (49.98) | (49.86) |
| Child Mortality | 81.71 | 70.16 | 84.85 | 73.77 | 91.30 | 81.64 | 98.11 | 89.99 | 90.68 | 82.56 | 86.75 | 73.98 |
| | (34.05) | (32.37) | (34.11) | (32.94) | (33.44) | (33.57) | (31.45) | (32.88) | (32.08) | (29.92) | (35.84) | (36.43) |
| Log(Avg Grain Price) | 0.0567 | -0.0506 | 0.0575 | -0.0439 | 0.0591 | -0.0267 | 0.0602 | -0.00625 | 0.0629 | -0.0409 | 0.0585 | -0.0163 |
| | (0.129) | (0.274) | (0.130) | (0.269) | (0.129) | (0.259) | (0.129) | (0.244) | (0.132) | (0.267) | (0.128) | (0.263) |
| Cases | 1,285,175 | 1,143,710 | 129,050 | 113,771 | 1,808,792 | 1,551,974 | 123,848 | 104,356 | 247,947 | 200,449 | 593,016 | 545,442 |

**Table 3.** Regression Analysis for the Effects of Early-Life Recession on Adults' Cause-Specific Mortality without Individual Controls and Interactions

| Hazard: Cause-Specific Death | Malignant Neoplasm (YOB) | Malignant Neoplasm (YOB+1) | Diabetes Mellitus (YOB) | Diabetes Mellitus (YOB+1) | Cardiovascular (YOB) | Cardiovascular (YOB+1) | Influenza and Pneumonia (YOB) | Influenza and Pneumonia (YOB+1) | Chronic Respiratory (YOB) | Chronic Respiratory (YOB+1) | All Other Diseases (YOB) | All Other Diseases (YOB+1) |
|---|---|---|---|---|---|---|---|---|---|---|---|---|
| | b/se | b/se | b/se | b/se | b/se | b/se | b/se | b/se | b/se | b/se | b/se | b/se |
| Business Cycle, Year of Birth | -3.158*** (0.013) | | -2.835*** (0.042) | | -2.586*** (0.009) | | -1.941*** (0.040) | | -3.722*** (0.033) | | -2.465*** (0.015) | |
| Business Cycle, Year of Birth + 1 | | -3.493*** (0.016) | | -3.258*** (0.053) | | -2.607*** (0.011) | | -1.910*** (0.050) | | -3.726*** (0.039) | | -3.101*** (0.019) |
| Age Dummies | Yes | Yes | Yes | Yes | Yes | Yes | Yes | Yes | Yes | Yes | Yes | Yes |
| Decade Dummies | Yes | Yes | Yes | Yes | Yes | Yes | Yes | Yes | Yes | Yes | Yes | Yes |
| Dust Bowl Period (TX, NM, OK, KS, CO) | Yes | Yes | Yes | Yes | Yes | Yes | Yes | Yes | Yes | Yes | Yes | Yes |
| 1918 Spanish Flu | Yes | Yes | Yes | Yes | Yes | Yes | Yes | Yes | Yes | Yes | Yes | Yes |
| Social Security Act Starts | Yes | Yes | Yes | Yes | Yes | Yes | Yes | Yes | Yes | Yes | Yes | Yes |
| Emergency Relief Act Starts | Yes | Yes | Yes | Yes | Yes | Yes | Yes | Yes | Yes | Yes | Yes | Yes |
| AFDC Act Starts | Yes | Yes | Yes | Yes | Yes | Yes | Yes | Yes | Yes | Yes | Yes | Yes |
| World War II Period | Yes | Yes | Yes | Yes | Yes | Yes | Yes | Yes | Yes | Yes | Yes | Yes |
| Cases | 11,007,755 | 11,007,755 | 11,007,755 | 11,007,755 | 11,007,755 | 11,007,755 | 11,007,755 | 11,007,755 | 11,007,755 | 11,007,755 | 11,007,755 | 11,007,755 |

*Notes. Stratified by sex, state of death, and race. Standard errors are clustered at state of birth. Data spans the years 1968-2004. 20 percent random sample is used. (\* p<0.1, \*\* p<0.05, \*\*\* p<0.01)*

**Table 4.** Regression Analysis for the Effects of Early-Life Recession on Adults' Cause-Specific Mortality without Individual Controls

| Hazard: Cause-Specific Death | Malignant Neoplasm (YOB) | Malignant Neoplasm (YOB+1) | Diabetes Mellitus (YOB) | Diabetes Mellitus (YOB+1) | Cardiovascular (YOB) | Cardiovascular (YOB+1) | Influenza and Pneumonia (YOB) | Influenza and Pneumonia (YOB+1) | Chronic Respiratory (YOB) | Chronic Respiratory (YOB+1) | All Other Diseases (YOB) | All Other Diseases (YOB+1) |
|---|---|---|---|---|---|---|---|---|---|---|---|---|
| | b/se | b/se | b/se | b/se | b/se | b/se | b/se | b/se | b/se | b/se | b/se | b/se |
| Business Cycle, Year of Birth | -2.451*** (0.054) | | -2.568*** (0.103) | | -3.553*** (0.069) | | -4.096*** (0.107) | | -0.541*** (0.044) | | -2.354*** (0.077) | |
| Business Cycle, Year of Birth + 1 | | -4.339*** (0.055) | | -4.309*** (0.093) | | -4.531*** (0.081) | | -4.212*** (0.105) | | -4.562*** (0.059) | | -4.079*** (0.064) |
| Individual Age and Marital Status Dummies | Yes | Yes | Yes | Yes | Yes | Yes | Yes | Yes | Yes | Yes | Yes | Yes |
| Policy, Incident, and Decade Dummies | Yes | Yes | Yes | Yes | Yes | Yes | Yes | Yes | Yes | Yes | Yes | Yes |
| Grain Prices ×Birth State | Yes | Yes | Yes | Yes | Yes | Yes | Yes | Yes | Yes | Yes | Yes | Yes |
| Child Mortality×Birth State | Yes | Yes | Yes | Yes | Yes | Yes | Yes | Yes | Yes | Yes | Yes | Yes |
| Cases | 6,995,351 | 6,995,351 | 6,995,351 | 6,995,351 | 6,995,351 | 6,995,351 | 6,995,351 | 6,995,351 | 6,995,351 | 6,995,351 | 6,995,351 | 6,995,351 |

*Notes. Stratified by sex, state of death, state of birth, and race. Standard errors are clustered at state of birth. Policy variables include: Emergency Relief Act of 1933, AFDC Act of 1940, Social Security Act of 1935 Incident dummies include: World War II Period, 1918 Spanish Flu, and Dust Bowl Period for states TX, NM, OK, KS, CO Data spans the years 1979-2004. 20 percent random sample is used. (\* p<0.1, \*\* p<0.05, \*\*\* p<0.01)*

**Table 5.** Regression Analysis for the Effects of Early-Life Recession on Adults' Cause-Specific Mortality, Full Specification

| Hazard: | Malignant Neoplasm | | Diabetes Mellitus | | Cardiovascular | | Influenza and Pneumonia | | Chronic Respiratory | | All Other Diseases | |
|---|---|---|---|---|---|---|---|---|---|---|---|---|
|  | (YOB) | (YOB+1) | (YOB) | (YOB+1) | (YOB) | (YOB+1) | (YOB) | (YOB+1) | (YOB) | (YOB+1) | (YOB) | (YOB+1) |
|  | b/se | b/se | b/se | b/se | b/se | b/se | b/se | b/se | b/se | b/se | b/se | b/se |
| Cause-Specific Death | -2.165*** |  | -2.346*** |  | -3.086*** |  | -3.701*** |  | -0.912*** |  | -2.082*** |  |
| Business Cycle, Year of Birth | (0.046) |  | (0.086) |  | (0.059) |  | (0.089) |  | (0.068) |  | (0.077) |  |
| Business Cycle, Year of Birth + 1 |  | -2.834*** |  | -2.808*** |  | -2.496*** |  | -2.259*** |  | -3.308*** |  | -2.338*** |
|  |  | (0.056) |  | (0.124) |  | (0.042) |  | (0.091) |  | (0.081) |  | (0.059) |
| Individual Dummies | Yes | Yes | Yes | Yes | Yes | Yes | Yes | Yes | Yes | Yes | Yes | Yes |
| Policy, Incident, and Decade Dummies | Yes | Yes | Yes | Yes | Yes | Yes | Yes | Yes | Yes | Yes | Yes | Yes |
| Grain Prices ×Birth State | Yes | Yes | Yes | Yes | Yes | Yes | Yes | Yes | Yes | Yes | Yes | Yes |
| Child Mortality×Birth State | Yes | Yes | Yes | Yes | Yes | Yes | Yes | Yes | Yes | Yes | Yes | Yes |
| Cases | 2,100,655 | 2,100,655 | 2,100,655 | 2,100,655 | 2,100,655 | 2,100,655 | 2,100,655 | 2,100,655 | 2,100,655 | 2,100,655 | 2,100,655 | 2,100,655 |

*Notes. Stratified by sex, state of death, state of birth, and race. Standard errors are clustered at state of birth. Individual Characteristics include: Marital Status (2 dummies), Education (3 dummies), Age (16 dummies), and Occupation (10 dummies). Policy variables include: Emergency Relief Act of 1933, AFDC Act of 1940, Social Security Act of 1935 Incident dummies include: World War II Period, 1918 Spanish Flu, and Dust Bowl Period for states TX, NM, OK, KS, CO Data spans the years 1989-1999. 20 percent random sample is used. (\*p<0.1, \*\*p<0.05, \*\*\*p<0.01)*

**Table 6.** Heterogeneity by Different Gender-Race Groups for Malignant Neoplasms and Cardiovascular Diseases

| Groups Based on: | Malignant Neoplasm | | | | Cardiovascular | | | |
|---|---|---|---|---|---|---|---|---|
| Gender-Race | White Male | White Female | Black Male | Black Female | White Male | White Female | Black Male | Black Female |
|  | b/se | b/se | b/se | b/se | b/se | b/se | b/se | b/se |
| Business Cycle, Year of Birth | -2.973*** | -2.678*** | -2.344*** | -2.010*** | -3.160*** | -3.109*** | -2.156*** | -2.659*** |
|  | (0.064) | (0.073) | (0.109) | (0.149) | (0.065) | (0.068) | (0.102) | (0.126) |
| Cases | 1,551,553 | 1,295,822 | 209,871 | 183,252 | 1,551,553 | 1,295,822 | 209,871 | 183,252 |

*Notes. Stratified by sex, state of death, state of birth, and race. Standard errors are clustered at state of birth. Individual Characteristics include: Marital Status (2 dummies), Education (3 dummies), Age (16 dummies), and Occupation (10 dummies). Policy variables include: Emergency Relief Act of 1933, AFDC Act of 1940, Social Security Act of 1935 Incident dummies include: World War II Period, 1918 Spanish Flu, and Dust Bowl Period for states TX, NM, OK, KS, CO Data spans the years 1989-1999. 20 percent random sample is used. (\*p<0.1, \*\*p<0.05, \*\*\*p<0.01)*

**Table 7.** Heterogeneity by Different Gender-Race Groups for Diabetes Mellitus and Influenza Pneumonia

| Groups Based on | Diabetes Mellitus | | | | Influenza Pneumonia | | | |
|---|---|---|---|---|---|---|---|---|
| Gender-Race | White Male | White Female | Black Male | Black Female | White Male | White Female | Black Male | Black Female |
|  | b/se | b/se | b/se | b/se | b/se | b/se | b/se | b/se |
| Business Cycle, Year of Birth | -2.772*** | -3.215*** | -1.741*** | -2.214*** | -3.297*** | -2.726*** | -1.997*** | -2.380*** |
|  | (0.158) | (0.147) | (0.260) | (0.244) | (0.148) | (0.125) | (0.332) | (0.248) |
| Cases | 1,551,553 | 1,295,822 | 209,871 | 183,252 | 1,551,553 | 1,295,822 | 209,871 | 183,252 |

*Notes. Stratified by sex, state of death, state of birth, and race. Standard errors are clustered at state of birth. Individual Characteristics include: Marital Status (2 dummies), Education (3 dummies), Age (16 dummies), and Occupation (10 dummies). Policy variables include: Emergency Relief Act of 1933, AFDC Act of 1940, Social Security Act of 1935 Incident dummies include: World War II Period, 1918 Spanish Flu, and Dust Bowl Period for states TX, NM, OK, KS, CO Data spans the years 1989-1999. 20 percent random sample is used. (\* p<0.1, \*\* p<0.05, \*\*\* p<0.01)*

**Table 8.** Heterogeneity by Different Gender-Race Groups for Chronic Respiratory and All Other Diseases

| Groups Based on: | Chronic Respiratory Diseases | | | | All Other Diseases | | | |
|---|---|---|---|---|---|---|---|---|
| Gender-Race | White Male | White Female | Black Male | Black Female | White Male | White Female | Black Male | Black Female |
|  | b/se | b/se | b/se | b/se | b/se | b/se | b/se | b/se |
| Business Cycle, Year of Birth | -3.999*** | -3.732*** | -3.206*** | -1.912*** | -1.908*** | -2.336*** | -0.726*** | -1.336*** |
|  | (0.076) | (0.092) | (0.302) | (0.312) | (0.113) | (0.082) | (0.184) | (0.135) |
| Cases | 1,551,553 | 1,295,822 | 209,871 | 183,252 | 1,551,553 | 1,295,822 | 209,871 | 183,252 |

*Notes. Stratified by sex, state of death, state of birth, and race. Standard errors are clustered at state of birth. Individual Characteristics include: Marital Status (2 dummies), Education (3 dummies), Age (16 dummies), and Occupation (10 dummies). Policy variables include: Emergency Relief Act of 1933, AFDC Act of 1940, Social Security Act of 1935 Incident dummies include: World War II Period, 1918 Spanish Flu, and Dust Bowl Period for states TX, NM, OK, KS, CO Data spans the years 1989-1999. 20 percent random sample is used. (\* p<0.1, \*\* p<0.05, \*\*\* p<0.01)*

**Table 9.** Heterogeneity by Cohorts for Malignant Neoplasms and Cardiovascular Diseases

| Hazard: | Malignant Neoplasm | | | | Cardiovascular | | | |
|---|---|---|---|---|---|---|---|---|
|  | 1910-1918 | 1919-1928 | 1929-1942 | 1943-1950 | 1910-1918 | 1919-1928 | 1929-1942 | 1943-1950 |
| Cause-Specific Death | b/se | b/se | b/se | b/se | b/se | b/se | b/se | b/se |
| Business Cycle, Year of Birth | -1.553*** | -15.136*** | -1.530*** | 5.302*** | -0.841*** | -15.322*** | -1.363*** | 5.445*** |
|  | (0.093) | (0.038) | (0.040) | (0.026) | (0.092) | (0.037) | (0.044) | (0.027) |
| Cases | 2,667,600 | 2,380,676 | 1,509,613 | 437,462 | 2,667,600 | 2,380,676 | 1,509,613 | 437,462 |

*Notes. Stratified by sex, state of death, state of birth, and race. Standard errors are clustered at state of birth. Individual Characteristics are restricted to age and marital status dummies. Data spans the years 1979-2004. 20 percent random sample is used. (\* p<0.1, \*\* p<0.05, \*\*\* p<0.01)*

**Table 10.** Heterogeneity by Cohorts for Diabetes Mellitus and Influenza Pneumonia

| Hazard: | Diabetes Mallitus | | | | Influenza Pneumonia | | | |
|---|---|---|---|---|---|---|---|---|
| | 1910-1918 | 1919-1928 | 1929-1942 | 1943-1950 | 1910-1918 | 1919-1928 | 1929-1942 | 1943-1950 |
| Cause-Specific Death | b/se | b/se | b/se | b/se | b/se | b/se | b/se | b/se |
| Business Cycle, Year of Birth | -0.730*** | -12.632*** | -0.658*** | 4.701*** | 0.289*** | -13.990*** | -0.648*** | 5.188*** |
| | (0.125) | (0.129) | (0.070) | (0.108) | (0.104) | (0.102) | (0.090) | (0.162) |
| Cases | 1,560,667 | 1,338,831 | 828,353 | 233,187 | 1,560,667 | 1,338,831 | 828,353 | 233,187 |

*Notes. Stratified by sex, state of death, state of birth, and race. Standard errors are clustered at state of birth. Individual Characteristics are restricted to age and marital status dummies. Data spans the years 1979-2004. 20 percent random sample is used. (\* p<0.1, \*\* p<0.05, \*\*\* p<0.01)*

**Table 11.** Heterogeneity by Cohorts for Chronic Respiratory and All Other Diseases

| Hazard: | Chronic Respiratory Diseases | | | | All Other Diseases | | | |
|---|---|---|---|---|---|---|---|---|
| | 1910-1918 | 1919-1928 | 1929-1942 | 1943-1950 | 1910-1918 | 1919-1928 | 1929-1942 | 1943-1950 |
| Cause-Specific Death | b/se | b/se | b/se | b/se | b/se | b/se | b/se | b/se |
| Business Cycle, Year of Birth | -0.770*** | -12.388*** | 0.433*** | 5.053*** | -0.241*** | -11.322*** | -1.051*** | 4.346*** |
| | (0.120) | (0.078) | (0.077) | (0.138) | (0.084) | (0.047) | (0.061) | (0.039) |
| Cases | 1,264,927 | 1,118,657 | 709,278 | 201,047 | 1,264,927 | 1,118,657 | 709,278 | 201,047 |

*Notes. Stratified by sex, state of death, state of birth, and race. Standard errors are clustered at state of birth. Individual Characteristics are restricted to age and marital status dummies. Data spans the years 1979-2004. 20 percent random sample is used. (\* p<0.1, \*\* p<0.05, \*\*\* p<0.01)*

**FIGURES**

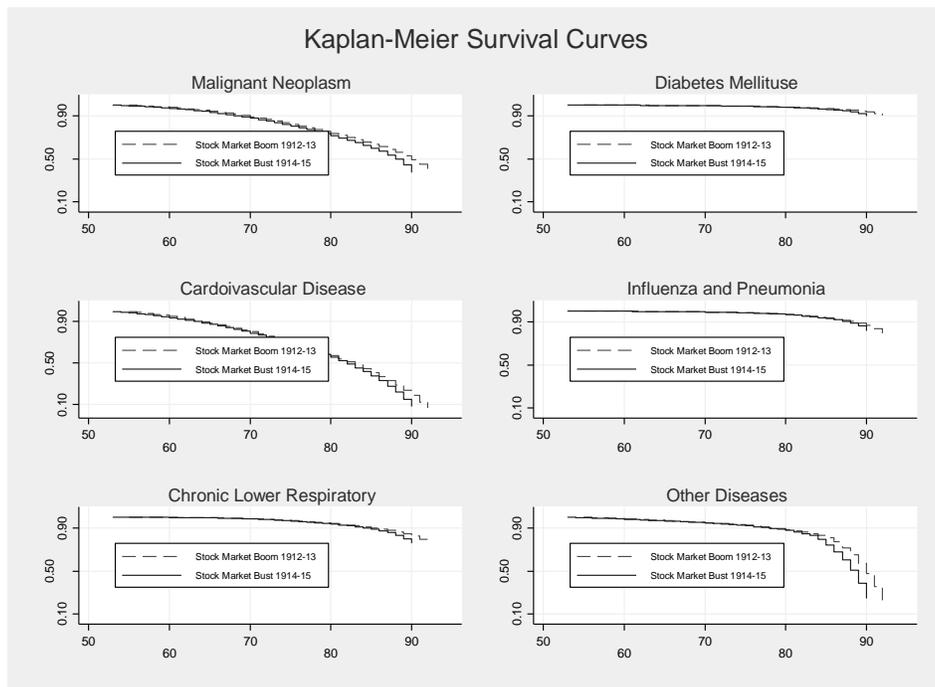

**Figure 1.** Mortality risks are constrained to males who died in period 1968-2004 and was born in the years 1912-13 (expansion) or 1914-15 (recession).

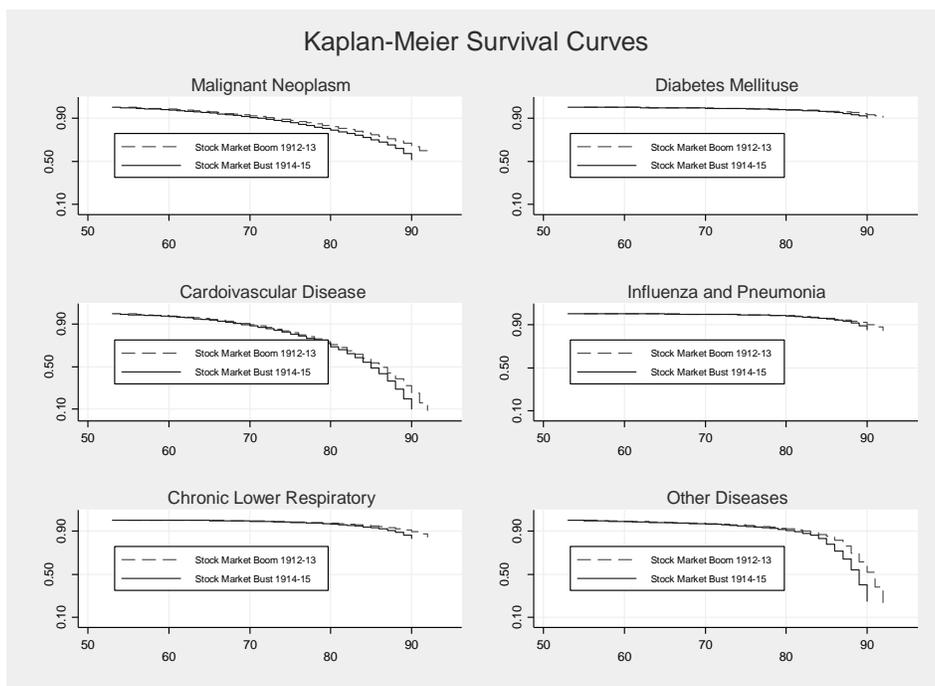

**Figure 2.** Mortality risks are constrained to females who died in period 1968-2004 and was born in the years 1912-13 (expansion) or 1914-15 (recession).

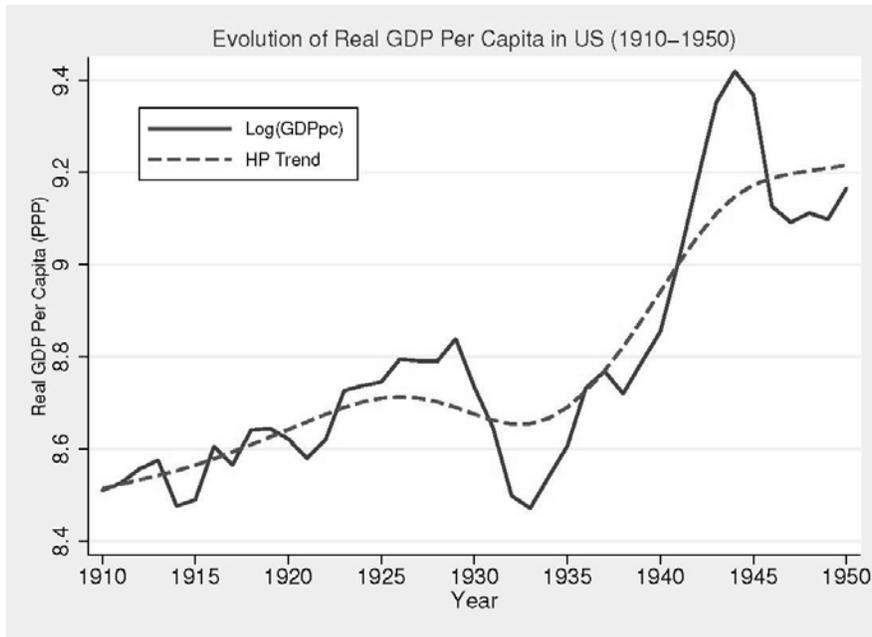

**Figure 3.** Log of GDP per capita (PPP) in United States during 1910-1950. The trend extracted using a Hodrick-Prescott filter with smoothing factor of 200.

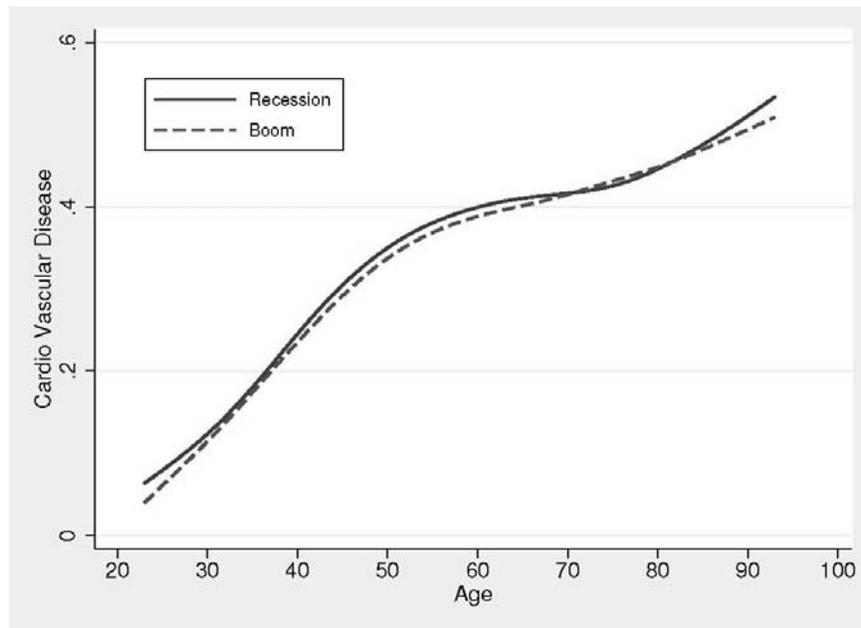

**Figure 4.** Likelihood of death from CVD for cohorts born during recession and expansions in different stages of life.